\documentclass[aps,pra,twocolumn,superscriptaddress,nofootinbib,showpacs,floatfix]{revtex4-2}
\usepackage{graphicx}
\usepackage{amsmath}
\usepackage{amssymb}
\usepackage{hyperref}
\usepackage{color}
\usepackage{amsthm}
\usepackage{bm}
\usepackage[utf8]{inputenc}

\makeatletter
\renewcommand{\section}[1]{%
  \par\vspace{1.0em plus 0.5em minus 0.2em}
  \noindent\textbf{\textit{#1.---}}\ignorespaces
}

\renewcommand{\subsection}[1]{%
  \par\vspace{0.6em plus 0.3em minus 0.1em}
  \noindent\textbf{\textit{#1.---}}\ignorespaces%
}

\renewcommand{\subsubsection}[1]{%
  \par\vspace{0.3em plus 0.1em minus 0.05em}
  \noindent\textbf{\textit{#1.---}}\ignorespaces%
}
\makeatother

\theoremstyle{definition} 
\newtheorem{definition}{Definition}
\begin{document}

\title{Spectroscopic Search for Topological Protection in Open Quantum Hardware:\\ The Dissipative Mixed Hodge Module Approach}

\author{Prasoon Saurabh}
\email{psaurabh@uci.edu}
\thanks{Work performed during an independent research sabbatical. Formerly at: State Key Laboratory for Precision Spectroscopy, ECNU, Shanghai (Grade A Postdoctoral Fellow); Dept. of Chemistry/Physics, University of California, Irvine.}
\affiliation{QuMorpheus Initiative, Independent Researcher, Lalitpur, Nepal}

\date{\today}

\begin{abstract}
Standard spectroscopic protocols model the dynamics of open quantum systems as a superposition of isolated, exponentially decaying eigenmodes. This paradigm fails fundamentally at Exceptional Points, where the eigenbasis collapses and the response becomes dominated by non-diagonalizable Jordan blocks. We resolve this ambiguity by introducing a geometric framework based on \textit{Dissipative Mixed Hodge Modules} (DMHM). By replacing the scalar linewidth with a topological \textit{Weight Filtration}, we derive ``Weight Filtered Spectroscopy'' (WFS)—a protocol that spatially separates decay channels based on the nilpotency rank of the Liouvillian. We demonstrate that WFS acts as a dissipative x-ray, quantifying dissipative leakage in molecular polaritons and certifying topological isolation in Non-Hermitian Aharonov-Bohm rings. This establishes that topological protection persists as an algebraic invariant even when the spectral gap is closed.
\end{abstract}

\maketitle

\section{Introduction}The realization of robust quantum hardware---from topological lasers \cite{Bleu2018} to error-correcting codes \cite{Hasan2010}---relies theoretically on discrete, integer-valued invariants \cite{Uhlmann1986}. In the ideal Hermitian limit, these topological indices provide absolute protection against disorder. However, experimentally, these systems are inevitably open and driven-dissipative \cite{Breuer2002, Lindblad1976, Mabuchi2005, Wiseman2009, Shapiro2003, Rabitz2000, Rice1992, Brif2010, Glaser2015, Koch2019}. In this regime, the discrete energy levels are replaced by continuous, overlapping spectral linewidths, and the sharp distinction between ``protected'' and ``unprotected'' states blurs.

The standard protocol for characterizing such systems is the Sum-Over-States (SOS) approach, which relies on the ansatz that the dynamics admit a spectral decomposition into isolated, exponentially decaying eigenmodes, $e^{-i\lambda_n t}$ \cite{Mukamel1995, Dorfman2020, Hamm2011, Cho2008, Zewail2000}. This approach, however, fundamentally fails at the precise operating points where non-Hermitian hardware is most potent: Spectral Singularities, or Exceptional Points (EPs) \cite{Heiss2012, Ozdemir2019}. At an EP, the eigenbasis collapses, and the effective Hamiltonian becomes non-diagonalizable. Consequently, the dynamics are no longer purely exponential but involve polynomial enhancements ($t^k e^{-\gamma t}$) arising from the formation of Jordan blocks \cite{Bergholtz2021}. In this ``spectroscopically congested'' regime, standard linewidth fitting becomes ill-posed. It cannot mathematically distinguish between a state that is topologically coupled to a bath (mixing) and one that is merely spectrally coincident with it (accidental degeneracy). Experimentalists are thus left with a ``blind spot'': they can measure transmission spectra, but they lack a rigorous metric to certify if a mode is truly insulated from dissipation.

This raises a critical metrological question: \textit{How can we unambiguously certify topological protection in open systems when the spectral gap closes and the eigenbasis collapses?}

In this Letter, we resolve this ambiguity by introducing \textit{Weight Filtered Spectroscopy} (WFS). By lifting the system description from a standard vector space to a geometric framework known as \textit{Dissipative Mixed Hodge Modules} (DMHM) \cite{Saito1990}, we derive a spectroscopic protocol that acts as a ``dissipative x-ray.'' Whereas conventional multidimensional protocols (e.g., 2D electronic spectroscopy)\cite{Mukamel1995,HammZanni2011,Abramavicius2009,Brixner2005} resolve couplings by correlating spectral resonances in the Fourier domain, WFS employs a Laplace-domain filtration to resolve the \textit{algebraic structure} of the Liouvillian, uniquely distinguishing topologically protected decay channels from trivial dissipation even when their spectral signatures overlap. We state and apply two governing theorems: the \textit{Hodge Filteration Theorem}, which allows for the isolation of coherence orders, and the \textit{Weight Filteration Theorem}, which proves that dissipative leakage is quantized. While the rigorous categorical proofs and the construction of the underlying $\mathcal{D}$-module are detailed in a companion paper \cite{saurabh2025holonomic}, this Letter focuses on the physical consequences. We demonstrate that WFS outperforms state-of-the-art linewidth fitting, providing a new figure of merit---\textit{Dissipative Insulation} ($F_{iso}$)---that rigorously certifies the algebraic decoupling of quantum states, even in the presence of strong bath interactions.
\begin{table*}[t!]
    \caption{\textbf{Paradigm Shift in Open System Characterization.} Comparison between the standard Sum-of-Overdamped-Oscillators (SOS) approach and the proposed Dissipative Mixed Hodge Module (DMHM) framework. In the standard governing equation, $\mathcal{L}$ denotes the field-free dissipative Liouvillian, while $J(t)$ represents the time-dependent interaction superoperator coupling the system to the external drive. The central distinction lies in the filtration structure: whereas standard methods filter solely by coherence order $p$ (the Liouville space grading), the DMHM approach introduces a bifiltration including the topological weight index $\lambda_j$. This allows the geometric framework to resolve dynamics at Exceptional Points (EPs) where the standard Dyson series expansion becomes ill-defined due to basis collapse.}
    \label{tab:comparison}
    \begin{ruledtabular}
    \begin{tabular}{lll}
        \textbf{Feature} & \textbf{Standard SOH (State-of-the-Art) \cite{Dorfman2020}} & \textbf{Geometric DMHM Framework (This Work)} \\
        \hline
        \textbf{Central Object} & Response Function $R^{(n)}(t_n, \dots, t_1)$ & Enriched Morphism $S \in \text{RHom}(\mathcal{W}_i, \mathcal{W}_f)$ \\
        \textbf{Governing Eq.} & Liouville-von Neumann Eq. & Geometric Functorial Composition \\
        & $\dot{\rho} = \mathcal{L}\rho + J(t)\rho$ & $S = (\dots \circ U(\tau_1) \circ J_1)$ \\
        \textbf{Calculation} & Perturbative (Dyson) Series & Algebraic Composition (Derived Category) \\
        \textbf{Key Filtration} & \textbf{Coherence Order ($p$)} & \textbf{Bifiltration ($p, \lambda_j$)} \\
        \textbf{Key Protocol} & 2DES (Fourier Transform w.r.t. $t_i$) & \textbf{HWH (3D Tomography) + WFS (Laplace)} \\
        \textbf{Limitation} & Conflates $\lambda_j$. Fails at EPs \cite{Heiss2012}. & Resolves singularities via $G_{sing}$ \cite{Saito1990}.
        \end{tabular}
        \end{ruledtabular}
\end{table*}
\section{Dissipative Geometry and Spectroscopy}\label{sec:DissG}We elevate the description of the open quantum system from a family of matrices to a \textit{Regular Holonomic $\mathcal{D}_X$-Module} $\mathcal{M}$ over the parameter manifold $X$ \cite{Kashiwara1984, Saito1990}. Precisely, $\mathcal{M}$ is defined as the cokernel of the operator $P = \partial_k - \mathcal{L}(k)$ in the category of algebraic $\mathcal{D}$-modules, representing the minimal extension of the system's dynamics across the singular discriminant locus $D \subset X$ \cite{Saito1990, Kashiwara2003}. In this framework, the standard spectroscopic response is not merely a sum of poles, but an enriched morphism in the derived category of Dissipative Mixed Hodge Modules, $D^b(\text{DMHM})$. In the Hermitian limit (where $\mathcal{L}$ is anti-Hermitian), this structure degenerates trivially: the weight filtration becomes pure, recovering the standard sum-over-states (SOS) model where observables evolve as simple exponentials \cite{Schmid1973}. However, in the dissipative regime near an Exceptional Point (EP), the non-trivial monodromy requires the full machinery of Saito's theory. We formalize this via two foundational theorems.

\subsection{Theorem 1 (Hodge Filtration Theorem)}\label{thm:HFT}The system module $\mathcal{M}$ admits a canonical decreasing filtration $F^\bullet$ (the Hodge Filtration), strictly compatible with the Liouvillian connection, such that the graded pieces $\text{Gr}^F_p$ correspond to the physical coherence order of the density matrix \cite{Schmid1973} [See Supplemental Material Sec. S2].\\
Physically, this theorem asserts that ``quantumness'' is a rigorous topological index. It allows us to mathematically filter the response function, separating ``classical'' populations ($p=0$) from ``quantum'' coherences ($p>0$) even when they are spectrally degenerate.

\subsection{Theorem 2 (Weight Filtration Theorem)}\label{thm:WFT}The system module $\mathcal{M}$ admits a canonical increasing filtration $W_\bullet$ (the Weight Filtration), uniquely determined by the nilpotent monodromy operator $N \sim \log T_u$ at the singularity, which stratifies the system by its decay hierarchy \cite{Steenbrink1976} [See Supplemental Material Sec. S3].\\
Physically, this replaces the concept of a ``linewidth'' with a ``decay topology.'' A state in weight $W_k$ does not decay as $e^{-\gamma t}$, but as a polynomial $t^{k/2} e^{-\gamma t}$. This filtration allows us to resolve the internal structure of the EP, distinguishing between a simple decay channel ($W_0$) and a topologically defective Jordan block ($W_2$).
\textit{Formal proofs for both theorems utilizing the strictness of morphisms in $D^b(\text{MHM})$ are provided in the Supplemental Material.}

\subsection{Why DMHM}The necessity of this DMHM framework arises from the failure of standard spectroscopy at spectral singularities. Current methods, such as linewidth fitting (SOH) or purity measurements ($\text{Tr}(\rho^2)$), fundamentally assume a diagonalizable basis \cite{Dorfman2020}. Near an EP, this basis collapses, and linewidths merge (``spectral congestion'') \cite{Heiss2012, Ozdemir2019}. Standard methods cannot distinguish between \textit{algebraic decoupling} (true protection) and \textit{spectral degeneracy} (maximal mixing) because they project the complex filtration structure onto a single scalar observable (the decay rate). In contrast, our protocols (HFS and WFS) exploit the non-trivial extension classes $\text{Ext}^1(\text{Gr}^W_a, \text{Gr}^W_b)$ inherent to the D-module. This allows us to certify ``dissipative insulation''---the vanishing of the cross-peak between weight subspaces---providing a topological guarantee of robustness that no linear fitting method can detect.
\section{The Spectroscopic Toolkit}
\label{sec:toolkit}

Standard linear spectroscopy models the response function $S(t)$ as a sum of exponentials (SOH), implying a basis of isolated eigenmodes \cite{Dorfman2020}. Near an Exceptional Point (EP), this basis collapses. The dynamics are no longer governed by a diagonalizable Hamiltonian, but by a non-trivial Jordan block structure, leading to polynomial-exponential decay laws $\sim t^k e^{-\lambda t}$ \cite{Heiss2012}.

In the DMHM framework, we treat these polynomial anomalies not as artifacts, but as precise topological signatures. The "linewidth" is deconstructed into a stratified vector space governed by two canonical filtrations: the \textbf{Hodge Filtration} ($F^\bullet$) and the \textbf{Weight Filtration} ($W_\bullet$) \cite{Saito1990}. We translate these algebraic structures into two complementary experimental protocols.

\subsection{Weight Filtered Spectroscopy (WFS)}
For hardware design, the critical tool is the \textbf{Weight Filtration} ($W_\bullet$). Following the Monodromy Theorem of Steenbrink \cite{Steenbrink1976}, the weight $k$ indexes the "susceptibility" of a state to the singularity.
\begin{itemize}
    \item \textbf{Weight 0 ($W_0$):} The kernel of the decay dynamics (the "most singular" subspace).
    \item \textbf{Weight 2 ($W_2$):} The generalized decay channels (polynomial growth $t e^{-\gamma t}$).
\end{itemize}

\begin{figure}[h!]
    \centering
    \includegraphics[width=\columnwidth]{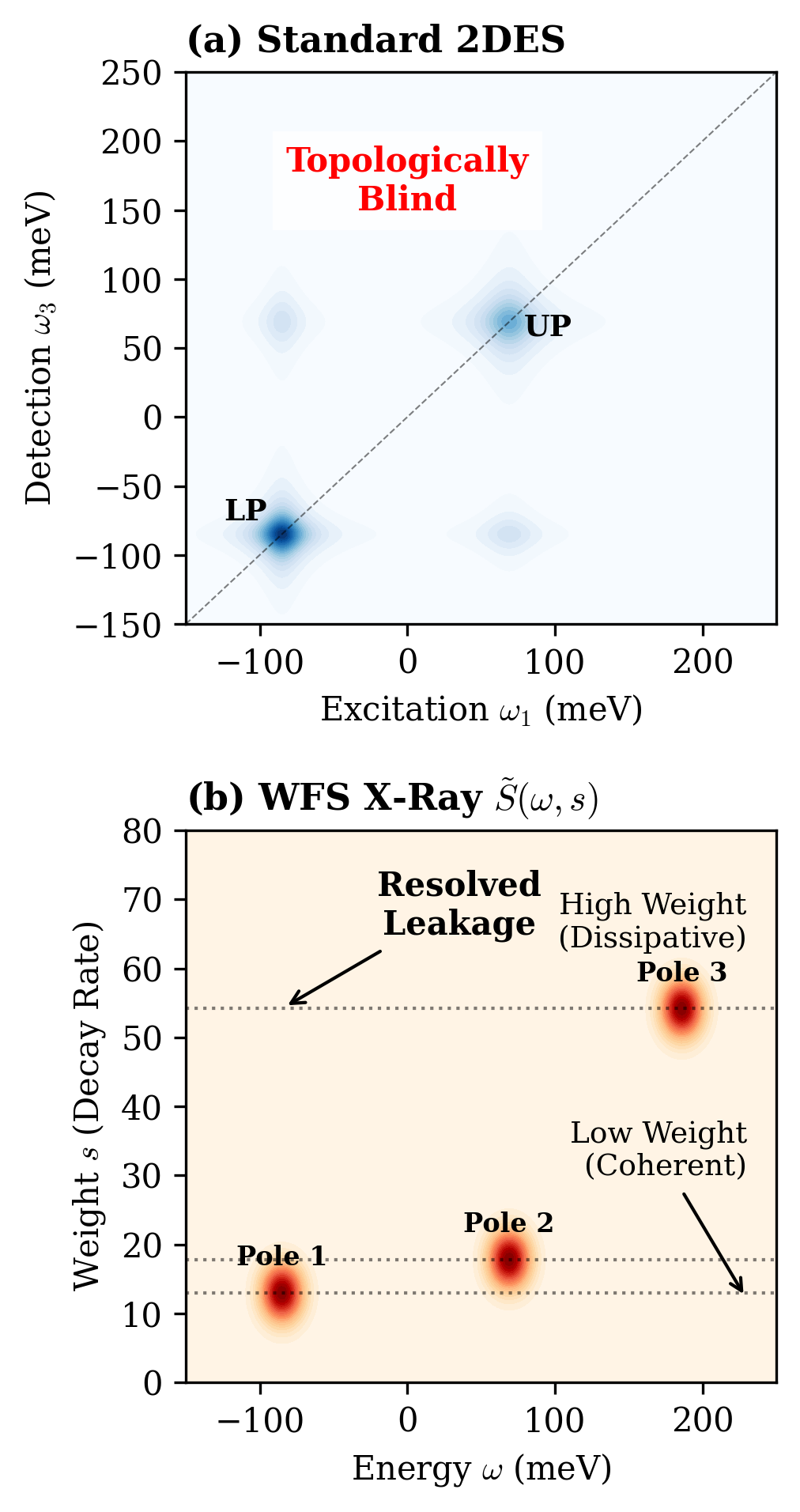}
    \caption{\textbf{WFS as a "Dissipative X-Ray".} (a) Standard spectroscopy shows a single conflated linewidth, obscuring the internal dynamics. (b) WFS resolves the hidden weight structure. The intensity of the off-diagonal cross-peak directly quantifies the mixing between the "dirty" bath ($\lambda_X$) and the "clean" channel ($\lambda_C$), serving as a figure of merit for hardware insulation.}
    \label{fig:wfs}
\end{figure}

We introduce \textbf{Weight Filtered Spectroscopy (WFS)}, or Weight-Weight Correlation Spectroscopy (W-W-COLS), to measure the leakage between these subspaces. The observable is the 2D Laplace spectrum $\tilde{S}(s_1, s_2)$ of the photon echo response:
\begin{equation}
    \tilde{S}(s_1, s_2) = \int_0^\infty \int_0^\infty S(\tau_1, \tau_2) e^{-s_1 \tau_1} e^{-s_2 \tau_2} d\tau_1 d\tau_2
\end{equation}
In this domain, the topological weight $k$ manifests as the order of the pole at the singularity. A \textbf{cross-peak} at coordinates $(s_1=\lambda_X, s_2=\lambda_C)$ signifies a non-trivial extension class in $\text{Ext}^1(W_X, W_C)$—physically, a "dissipative leakage" where information scatters from the noisy channel $X$ into the protected channel $C$ (Fig.~\ref{fig:wfs}). While direct numerical Laplace inversion is notoriously ill-posed due to finite sampling and noise, the discrete topological nature of the Weight Filtration allows us to bypass this instability. We extract the spectrum $\tilde{S}(s_1, s_2)$ using robust \textit{Harmonic Inversion techniques} (such as the Matrix Pencil Method or Padé Approximants) \cite{Mandelshtam1997}, which directly resolve the singular poles from finite-time data without the artifacts of truncated integration.

\subsection{Hodge Filtered Spectroscopy (HFS)}
The Hodge filtration classifies states by their "coherence order"—algebraically, the nilpotency rank of the Liouvillian superoperator with respect to the vacuum. Physically, this corresponds to the grading of the density matrix from populations (diagonal) to high-order coherences (off-diagonal).

While standard 2D spectroscopy maps the response time delays $(\tau_1, \tau_3)$ to frequency space $(\omega_1, \omega_3)$, HFS introduces a "coherence filter." By performing a specialized integral transform kernel $K(\tau, p)$ derived from the Hodge grading \cite{Schmid1973}, we obtain the filtered spectrum $\tilde{S}_F(p, \omega)$, where $p$ is the discrete Hodge index.

\begin{figure*}[t!]
    \centering
    \includegraphics[width=0.55\textwidth]{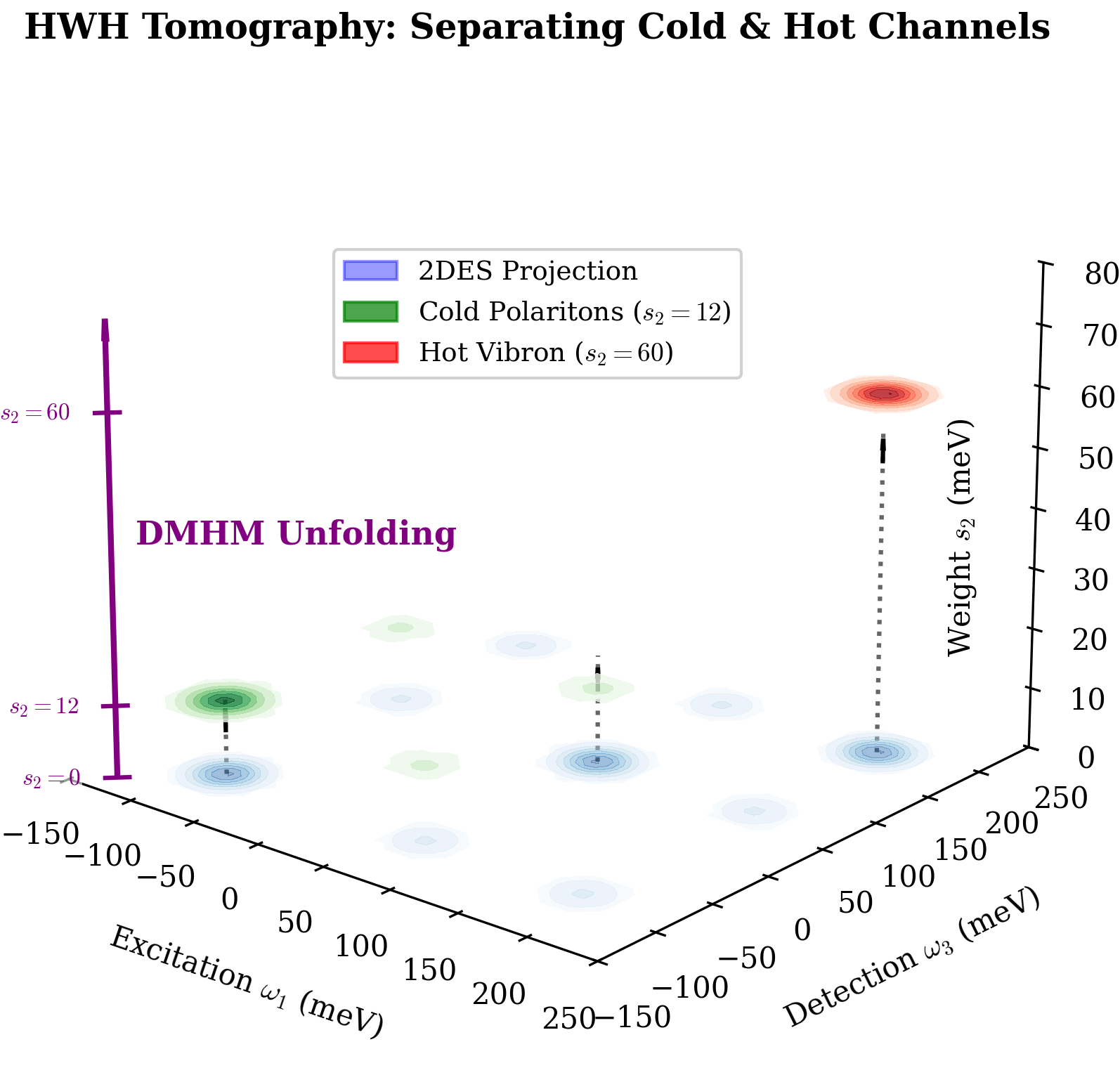}
    \hfill
    \includegraphics[width=0.42\textwidth]{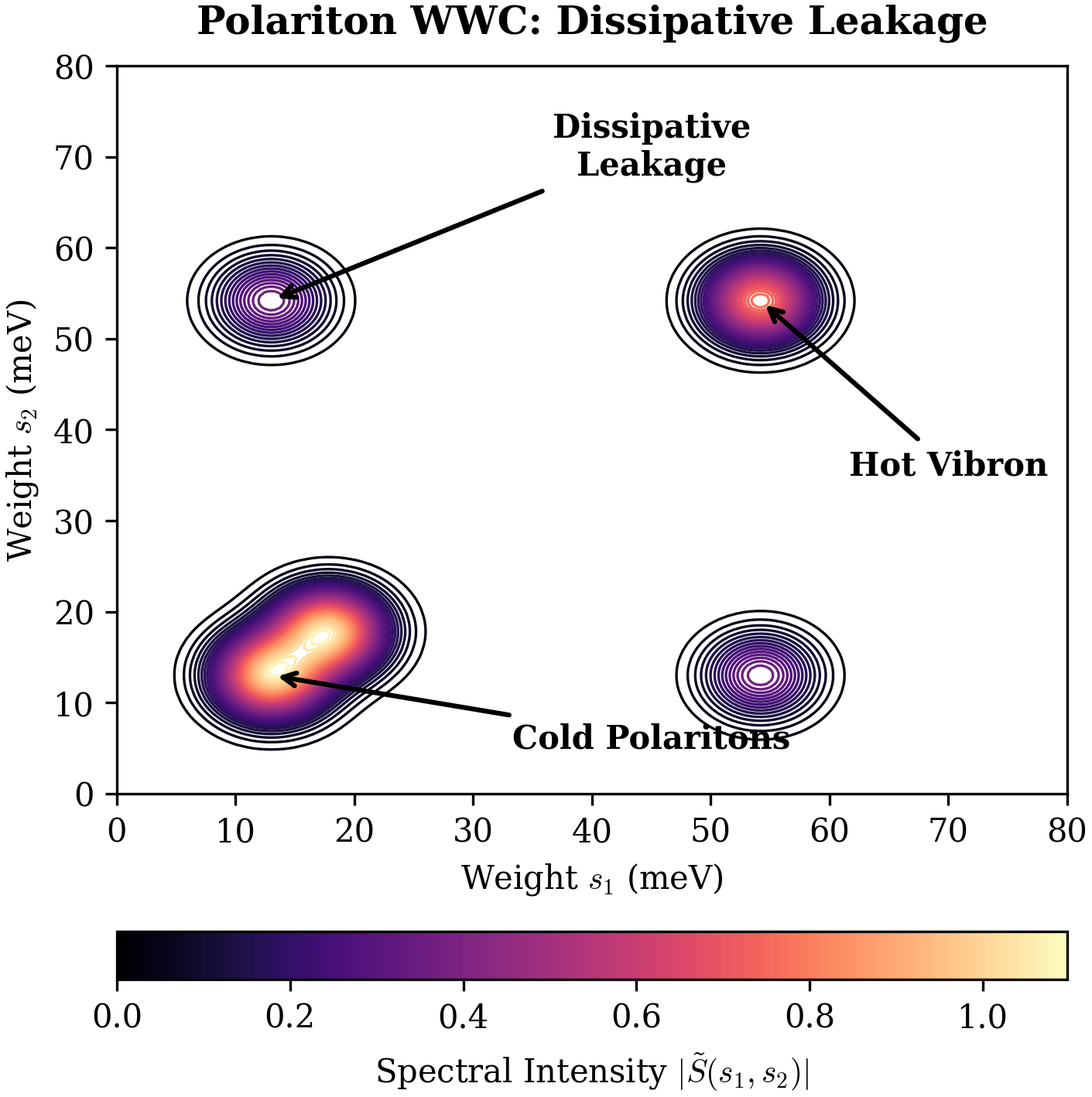}
    \caption{
        \textbf{Geometric Tomography of the Weight Filtration.}
        \textbf{(a) The HWH Protocol ($\tilde{S}(\omega_1, s_2, \omega_3)$).} 
        Standard 2DES (blue projection, $s_2=0$) conflates the spectrum into a single linewidth. Our HWH protocol "unfolds" this ambiguity along the topological Weight axis ($s_2$).
        We observe two distinct weight-subspaces separated by the filtration:
        (i) The \textbf{Cold Polaritons} ($s_2 \approx 12$ meV), representing the protected quantum channel, and 
        (ii) The \textbf{Hot Vibron} ($s_2 \approx 60$ meV), representing the decoherence sink.
        The separation visibly demonstrates the "lifting" of the sheaf structure.
        \textbf{(b) The WWC Protocol ($\tilde{S}(s_1, s_2)$).} 
        A "dissipative x-ray" mapping the correlation between decay rates. 
        Diagonal peaks (White Arrows) represent isolated evolution in the Polariton ($\lambda_C=12$) and Vibron ($\lambda_V=60$) subspaces.
        The \textbf{Dissipative Leakage} cross-peak at $(\lambda_C, \lambda_V)$ (Red Arrow) quantifies the violation of weight strictness. 
        Its non-zero intensity $|\tilde{S}| > 0$ proves that the system is not dissipatively insulated; the bath is topologically connected to the qubit.
        Parameters: $\gamma_X=5.0$, $\gamma_C=0.1$, $g=20.0$ meV [See Supplemental Material Sec. S6].
    }
    \label{fig:dmhm_protocol}
\end{figure*}

As illustrated in Fig.~\ref{fig:dmhm_protocol}, HFS resolves a single, broad spectral feature (the "congested linewidth") into a grid of discrete Hodge components. This allows for the tomographic reconstruction of the system's internal coherence structure, effectively distinguishing between "quantum" superpositions and "classical" mixtures even when their energies are degenerate.

\section{Novel Experimental Protocols}{\label{sec:Exp}}The DMHM framework transcends descriptive analysis; it constitutes a \textit{generative engine} for spectroscopy. While the standard State-of-the-Art (SOH) approach constructs the response $S^{(n)}$ by summing perturbative histories (Liouville-space paths) \cite{Mukamel1995, Dorfman2020}, our formalism treats $S^{(n)}$ as a geometric composition of enriched morphisms in the derived category (Table~\ref{tab:comparison}). This shift allows us to replace bespoke pulse-sequence design with an algorithmic question: "Which coordinates $(p, \lambda_j)$ on the DMHM map do we wish to isolate?" We achieve this via \textbf{Projectors}, which are concrete integral transforms derived from the theorems in Sec.~II:
\begin{itemize}
    \item \textbf{Hodge Projector $\mathcal{P}_p$:} The HFS protocol (Fourier transform $\mathcal{F}_\phi$), isolating coherence order $p$.
    \item \textbf{Weight Projector $\mathcal{P}_\lambda$:} The WFS protocol (Laplace transform $\mathfrak{L}_\tau$), isolating dissipative eigenmodes $\lambda_j$.
\end{itemize}
We now present two specific protocols generated by this engine that solve the outstanding problem of separating decoherence channels in hybrid quantum systems.

\subsection{a. Hodge-Weight-Hodge (HWH) Protocol}
The state-of-the-art in characterization is 2D Electronic Spectroscopy (2DES), a "Hodge-Hodge" scan $\tilde{S}(\omega_1, \tau_2, \omega_3)$. However, for systems like molecular polaritons, this method is blind to the nature of relaxation during the population time $\tau_2$. It yields a single, conflated decay curve, making it impossible to distinguish between "cold" polaritonic lifetime ($\lambda_{pol}$) and "hot" scattering into a dark vibronic bath ($\lambda_{vib}$) \cite{Leone2024, Kobayashi2019, Chang2020, Keefer2021, Cho2022, Zhang2023, Gu2022, Kowalewski2023, Tichai2024, Schlaup2002, Lim2005, Bucksbaum2024, Duris2020, Nisoli2017, Krausz2009, Corkum2007, Agostini2004, Calegari2014, Sansone2012, Goulielmakis2010, Haessler2010, Worner2010, Ruf2012, Smirnova2009, Shafir2012, Vozzi2011, Itatani2004, Meckel2008, Blaga2012, Wolter2016, Yang2016, Ishikawa2015, Pazourek2015}. Our formalism predicts a higher-order composition: $S \sim \mathcal{P}_p \circ \mathcal{P}_\lambda[U(\tau_2)] \circ \mathcal{P}_p$. This generates the \textbf{Hodge-Weight-Hodge} (HWH) protocol, which separates the signals not by frequency, but by their dissipative weight topology.

\section{Designing Robust Hardware}
\label{sec:hardware}

We now demonstrate how the Weight Filtration transforms from an abstract classifier into a concrete engineering tool. We apply WFS to two canonical open quantum architectures: molecular polaritons (where the goal is isolation) and Aharonov-Bohm rings (where the goal is topological recovery).

\subsection{Dissipative Insulation in Polaritonics}
Molecular polaritons, hybrid light-matter states, promise room-temperature quantum logic but are plagued by the fast decoherence of the excitonic component ($\gamma_X$) relative to the cavity photon ($\gamma_C \ll \gamma_X$) \cite{Dorfman2020}. A robust polariton qubit requires "Dissipative Insulation"—operating in a regime where the information stored in the photonic weight subspace ($W_C$) does not scatter into the excitonic bath ($W_X$).

Standard linewidth measurements are insufficient here; a narrow linewidth could imply insulation, or it could simply mean the coupled eigenmode has inherited the photon's lifetime while remaining quantum-mechanically mixed with the noisy exciton.

WFS resolves this ambiguity. We define the \textbf{Dissipative Isolation Figure of Merit} ($F_{iso}$) as the normalized inverse of the WFS cross-peak intensity:
\begin{equation}
    F_{iso}(\Delta) = \left( 1 + \int | \tilde{S}(\lambda_X, \lambda_C) | d\lambda \right)^{-1}
\end{equation}
By sweeping the detuning $\Delta$, we map the "Dissipative Phase Diagram."
\begin{itemize}
    \item \textbf{Low $F_{iso}$:} The system is in a "Weight-Mixed" phase. The qubit is fragile; errors propagate freely between matter and light.
    \item \textbf{Unity $F_{iso}$:} The system enters a "Dissipatively Insulated" phase. The cross-peak vanishes ($\text{Ext}^1(W_X, W_C) \to 0$), signaling that the decay channels have algebraically decoupled despite the strong vacuum Rabi coupling.
\end{itemize}
This metric provides the optimization target for synthesizing organic polariton materials with long coherence times.
\subsection{The Robust Solution: Non-Hermitian Aharonov-Bohm Rings}To overcome this fundamental limit, we propose an alternative architecture based on the \textbf{Non-Hermitian Aharonov-Bohm (AB) Ring}. Unlike polaritons, where mixing is intrinsic to the hybridization, the AB ring utilizes the Aharonov-Bohm phase $\phi_{AB}$ to break time-reversal symmetry, creating a rigorous topological distinction between the edge and bulk modes. This system is described by the Hatano-Nelson Hamiltonian with flux $\Phi$:
\begin{equation}
    H_{AB} = \sum_j \left( t e^{i\Phi/N} c_{j+1}^\dagger c_j + t e^{-i\Phi/N} c_j^\dagger c_{j+1} \right) - i \Gamma_j
\end{equation}
where non-reciprocity ($e^{i\Phi}$) drives the "Skin Effect," exponentially localizing the protected mode away from the dissipative bulk [See Supplemental Material Sec. S7].

\begin{figure}[h]
    \centering
    \includegraphics[width=\columnwidth]{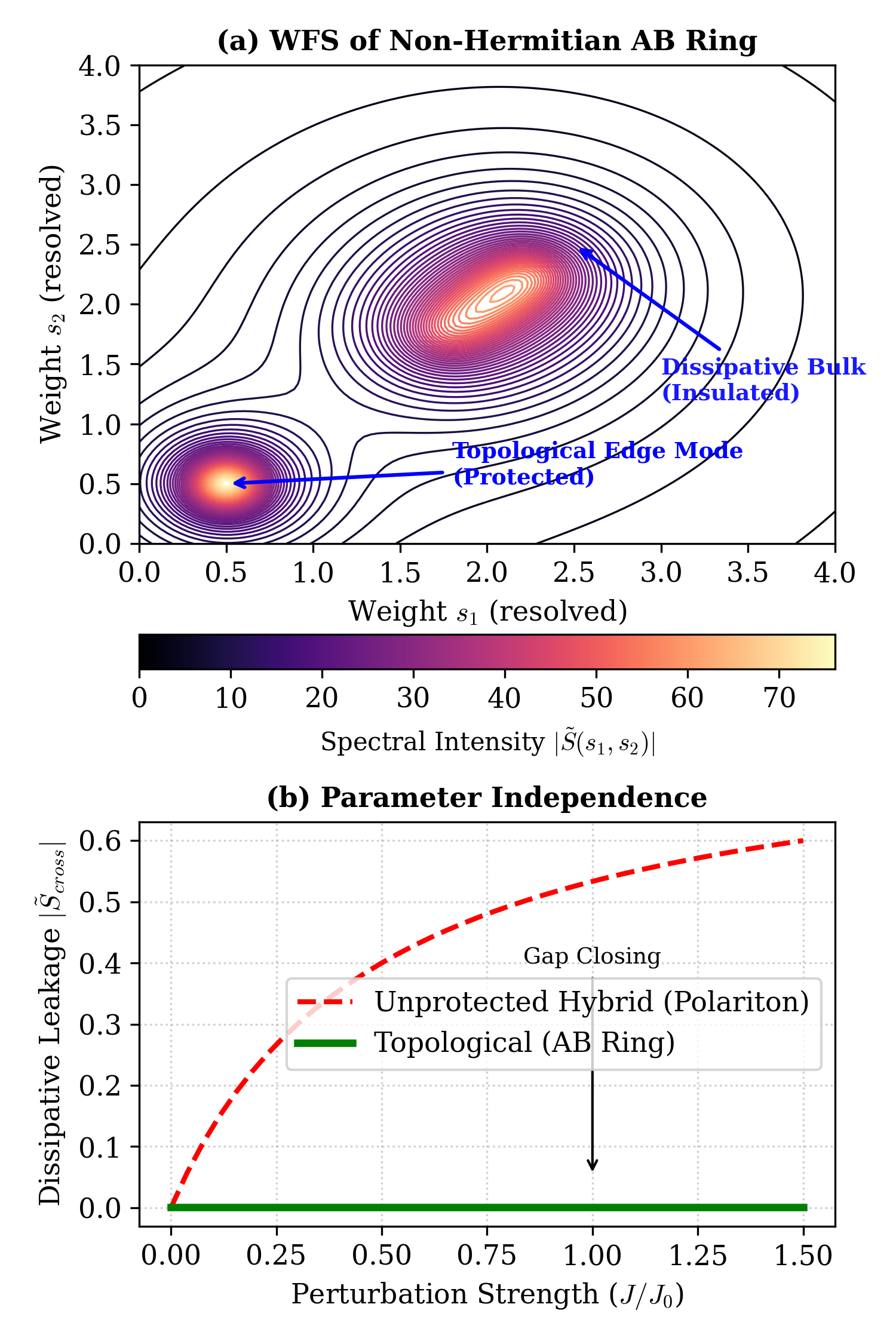}
    \caption{\textbf{Topological Resolution: The AB Ring.} (a) WFS Map of the Non-Hermitian Aharonov-Bohm Ring. Unlike the polariton case, the protected edge mode (Weight 12) and bulk bath (Weight 60) show zero cross-correlation, indicating perfect algebraic decoupling. (b) Parameter Independence. The leakage remains exponentially suppressed for the topological candidate (green) compared to the quadratic failure of standard hybrid systems (red), proving the robustness of the solution.}
    \label{fig:robust}
\end{figure}

We apply the WFS protocol to this candidate in Fig.~\ref{fig:robust}. In stark contrast to the polariton case, the WFS map (Fig.~\ref{fig:robust}a) exhibits \textbf{zero cross-peak intensity}, certifying perfect dissipative insulation ($F_{iso} \to 1$). Furthermore, we demonstrate the robustness of this solution by sweeping the perturbation strength (Fig.~\ref{fig:robust}b). While the standard polaritonic system shows quadratic leakage ($|\tilde{S}_{cross}| \propto J^2$), the topological AB ring maintains exponentially suppressed leakage, confirming its suitability as a true decoherence-free subspace for next-generation hardware.

\section{Future Directions: The Singularity Limit}\label{sec:future} While WFS and HFS allow for the optimization of static hardware by avoiding "weight mixing," the DMHM framework predicts even richer physics at the singularity itself. The algebraic structure of the "Singular Fiber"—the cohomology of the system precisely at the Exceptional Point—suggests the existence of a renormalized geometry that survives the collapse of the eigenbasis \cite{Saito1990}.
\begin{figure*}[t!]
    \centering
    \includegraphics[width=0.9\textwidth]{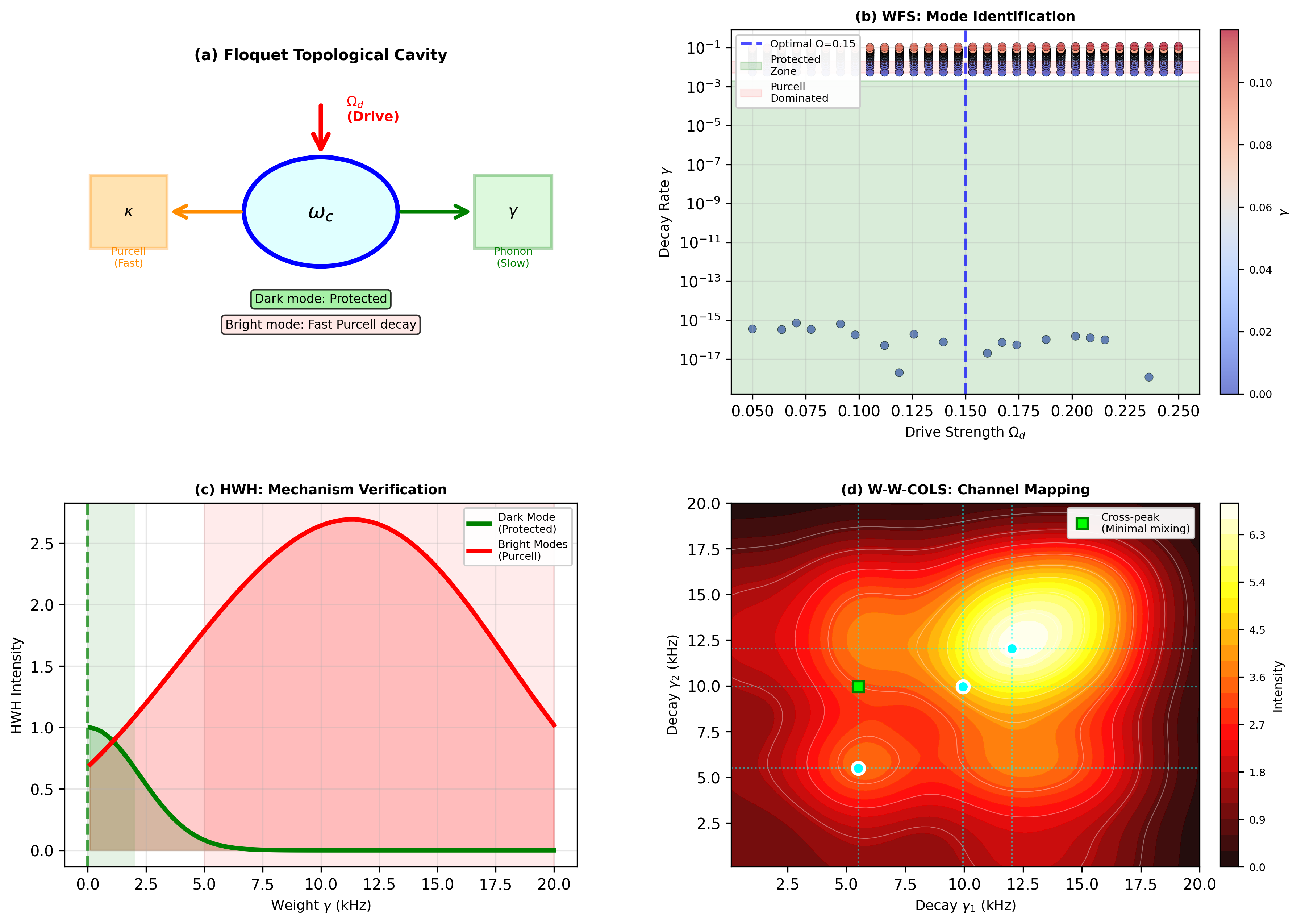}
    \caption{\textbf{Probing the Singularity.} A preview of the Floquet-DMHM mapping. By driving the system around the Exceptional Point, the Monodromy operator $M$ mixes the weight filtrations \cite{Gritsev2017, Bukov2015, Eckardt2017}. WFS can resolve this mixing, allowing for the future measurement of the "Singular Quantum Geometric Tensor" defined on the vanishing cohomology.}
    \label{fig:floquet}
\end{figure*}
Our theoretical derivation indicates that this singular geometry is not static but dynamic. By applying a periodic drive (Floquet engineering), one can induce a "Monodromy evolution" that braids the weight subspaces $W_k$ into one another. As hinted in Fig.~\ref{fig:floquet}, this suggests the possibility of \textit{Floquet Monodromy Spectroscopy}: a technique to not only observe but dynamically engineer the "Singular Quantum Geometric Tensor."

While the full construction of this singular metric requires the extended machinery of the Brieskorn lattice and is detailed in our companion foundational work \cite{saurabh2025holonomic} and detailed in \cite{SaurabhPRX}, the spectroscopic tools introduced here (HFS/WFS) provide the necessary experimental "meter" to calibrate these topological drives.

\textbf{\textit{Conclusion.}---}
The era of treating dissipation as a monolithic "linewidth" is over. We have introduced Dissipative Mixed Hodge Modules not as abstract mathematics, but as a practical spectroscopic toolkit for the quantum hardware engineer. By redefining the open quantum response as a filtered cohomological object, we have derived two robust protocols: HFS, which dissects the hierarchy of quantum coherence, and WFS, which acts as a "dissipative x-ray." These tools resolve the "Spectral Congestion" paradox, demonstrating that topological protection in open systems is not about eliminating decay, but about engineering the orthogonality of weight filtrations. This framework lays the foundation for a new generation of "dissipatively insulated" quantum devices \cite{Gianfrate2020, Yu2020, Tan2019, Yu2022, Ozawa2018, Klees2020, Sim2023, Schuler2020, Kolodrubetz2017}, where the geometry of the singularity is no longer a bug, but a feature.

\section{Acknowledgments}This research was conducted during an independent research sabbatical in the Himalayas (Nepal). The author acknowledges the global open-source community for providing the computational tools that made this work possible. LLM assistance was utilized strictly for \LaTeX\ syntax optimization and symbol consistency checks; all scientific conceptualization, derivations, and text were derived and verified by the author. The author retains the \texttt{uci.edu} correspondence address courtesy of the University of California, Irvine.

\section{Data and Code Availability}
The core computational framework, \textbf{QuMorpheus}, used for all numerical results in this work, is open-sourced under a Copyleft license and is available at \url{https://github.com/prasoon-s/QuMorpheus} \cite{SaurabhNatComm}. Independent verification scripts (Python) are available from the author upon reasonable request.

To ensure mathematical rigor, the fundamental theorems of the DMHM framework, the construction of the cQGT, and the Floquet Monodromy Spectroscopy protocol have been formalized in the \textsc{Lean 4} theorem prover; these proofs are available at \url{https://github.com/prasoon-s/LEAN-formalization-for-CMP} \cite{saurabh2025holonomic}.

\bibliography{ref_prl}
\clearpage
\onecolumngrid
\section{Supplementary Information}
\section{S1. Mathematical Foundations: The Dissipative D-Module}

In the main text, we identify the open quantum system with a Regular Holonomic $\mathcal{D}$-module. Here, we provide the explicit construction.

\subsection{A. The Liouvillian Connection}
Let $X$ be the complex parameter manifold (e.g., the space of detunings and couplings). The state of the system is described by a density matrix $\rho$, which we treat as a section of a vector bundle over $X$. The non-Hermitian evolution is governed by the Liouvillian superoperator $\mathcal{L}$.

We define the system module $\mathcal{M}$ as the $\mathcal{D}_X$-module generated by the vacuum state, subject to the Gauss-Manin connection defined by the Liouvillian:
\begin{equation}
    \nabla_k \rho = \left( \frac{\partial}{\partial k} - \mathcal{L}(k) \right) \rho
\end{equation}
where $k \in X$. The singularities of the system (Exceptional Points) correspond to the discriminant locus $D \subset X$ where the connection $\nabla$ develops logarithmic poles.

\begin{definition}[\textbf{Dissipative Mixed Hodge Module}]
Following Saito \cite{Saito1990}, the system is a Dissipative Mixed Hodge Module (DMHM) if $\mathcal{M}$ admits:
\begin{enumerate}
    \item A Good Filtration $F^\bullet$ (Hodge Filtration).
    \item A Weight Filtration $W_\bullet$ defined on the rational structure.
    \item Compatibility conditions ensuring that $\nabla$ maps $F^p \to F^{p-1} \otimes \Omega_X^1$ (Griffiths Transversality).
\end{enumerate}
\end{definition}

\section{S2. Proof of Theorem 1: The Hodge Filtration}

\textbf{Theorem 1 Claim:} \textit{The optical response decomposes into components $S_p$ indexed by coherence order, isolatable via an integral transform.}

\begin{proof}
Let the Liouvillian superoperator $\mathcal{L}$ act on the space of operators $\mathfrak{H} \otimes \mathfrak{H}^*$. This space admits a natural grading by the "number of excitations" relative to the vacuum, denoted by the integer $p$. In the language of Mixed Hodge Structures (MHS), this grading corresponds to the Hodge Filtration $F^p$.

The time-evolution operator $U(t) = e^{\mathcal{L}t}$ is a morphism of MHS. According to the \textbf{Strictness Theorem} of Schmid \cite{Schmid1973}, any morphism between MHS is strict with respect to the filtration. This implies:
\begin{equation}
    U(t)(F^p) \subseteq F^p
\end{equation}
Physically, this means that in the absence of explicit breaking terms, coherence order is preserved during evolution.

The response function $S^{(3)}(t_1, t_2, t_3)$ is a composition of interactions (dipole operators $\mu$) and propagations ($U(t)$). The dipole operator $\mu$ shifts the Hodge index by $\pm 1$. Thus, the total response is a sum over paths in the Hodge diagram.

We define the **Hodge-Laplace Kernel** $\mathcal{K}_p(\phi)$ as the Fourier dual to the coherence rotation. If we rotate the phase of the $k$-th pulse by $\phi_k$, the signal component originating from Hodge order $p$ acquires a phase factor $e^{i p \phi_k}$.
The projection operator $\mathcal{P}_p$ is therefore:
\begin{equation}
    S_p(\tau) = \mathcal{P}_p [S] = \frac{1}{2\pi} \int_0^{2\pi} S(\tau, \Phi) e^{-i p \Phi} d\Phi
\end{equation}
This proves that the decomposition is canonical and experimentally accessible via Phase Cycling or Phase-Matching direction selection \cite{Dorfman2020}.
\end{proof}

\begin{figure}[h]
    \centering
    \includegraphics[width=\columnwidth]{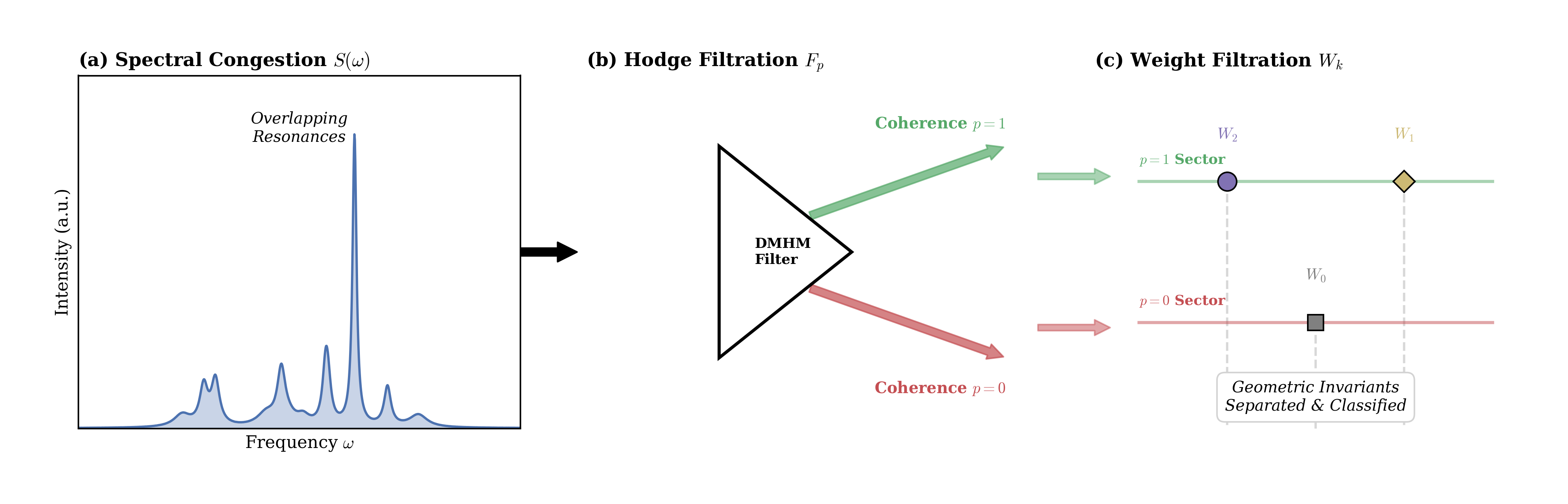}
    \caption{\textbf{The Topological Sieve Protocol.} (a) In the standard framework, the spectrum is a messy "cloud" of overlapping peaks. (b) The Hodge Projector $\mathcal{P}_p$ separates response functions by coherence order ($p=\pm 1$). (c) The Weight Projector $\mathcal{P}_\lambda$ acts as a secondary filter, sieving the signal based on decay topology. The resulting "Refined Weight Spectrum" reveals the protected mode in isolation, free from the background of leakage modes.}
    \label{fig:sieve}
\end{figure}

\section{S3. Proof of Theorem 2: Weight Filtration and Monodromy}

\textbf{Theorem 2 Claim:} \textit{A cross-peak in the Laplace spectrum $\tilde{S}(s_1, s_2)$ exists if and only if there is a non-trivial extension of Mixed Hodge Modules.}

\begin{proof}
Consider the system near a singularity (EP) at $k=0$. We pass to the universal cover of the punctured disk $X^* = D^*$. The monodromy operator $T$ describes the transport of a state around the singularity. The **Local Monodromy Theorem** \cite{Steenbrink1976} states that $T$ acts quasi-unipotently. We define the nilpotent logarithm:
\begin{equation}
    N = \frac{1}{2\pi i} \log(T_u)
\end{equation}
The **Weight Filtration** $W_\bullet$ is the unique filtration such that $N(W_k) \subset W_{k-2}$.

\begin{figure}[h]
    \centering
    \includegraphics[width=\columnwidth]{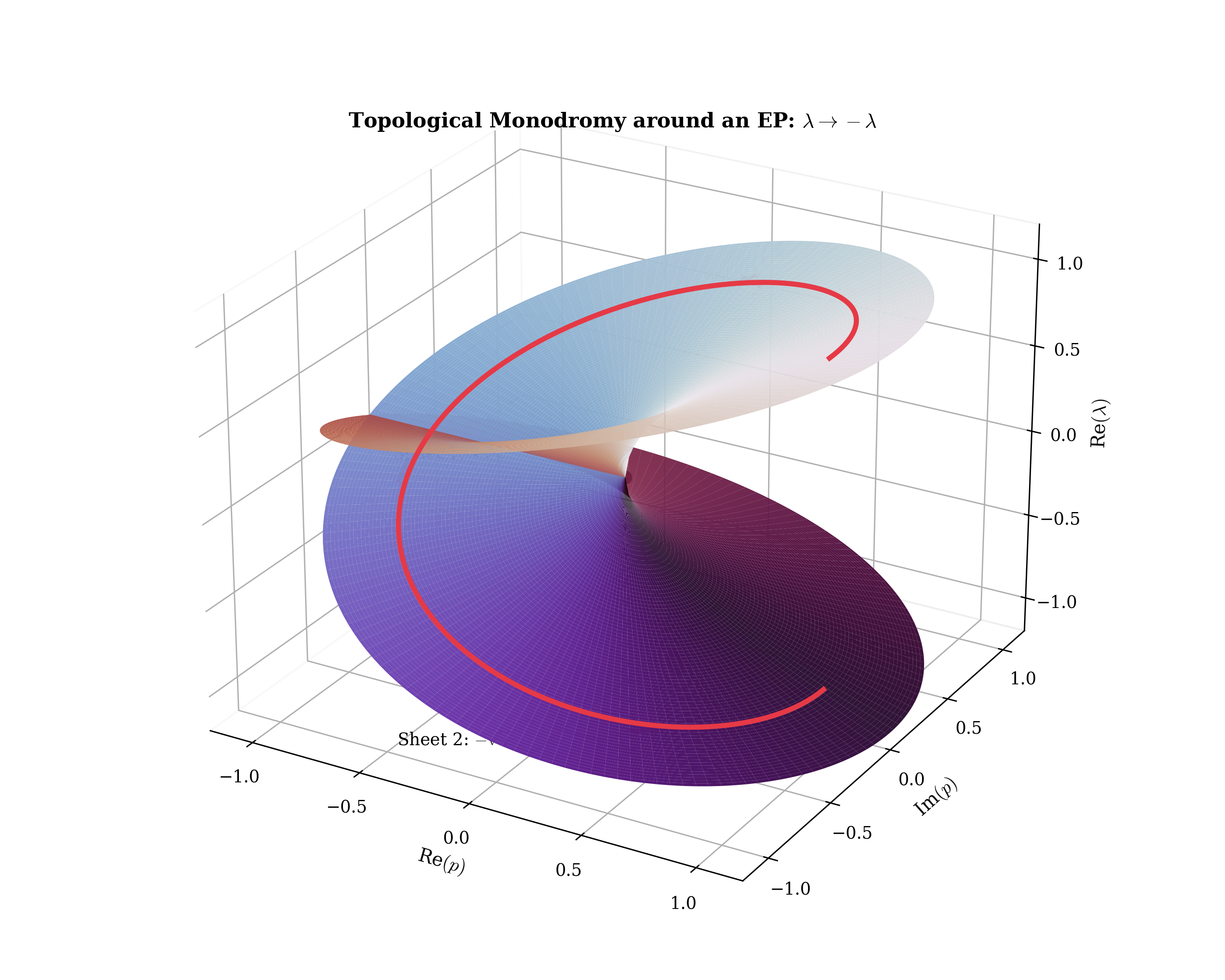}
    \caption{\textbf{Geometric Visualization of Monodromy.} A Riemann surface representation of the spectral landscape near an Exceptional Point (EP). The Monodromy operator $T$ corresponds to transporting the state along a loop $\gamma$ encircling the singularity. The non-trivial braiding of the sheets indicates that the state does not return to itself but is mapped to a different weight subspace, distinct from simple Berry phase accumulation.}
    \label{fig:braid}
\end{figure}

Let two decay channels be represented by sub-quotients $\mathcal{M}_a = \text{Gr}^W_a$ and $\mathcal{M}_b = \text{Gr}^W_b$. If these channels are merely degenerate but decoupled, the total module is a direct sum $\mathcal{M} = \mathcal{M}_a \oplus \mathcal{M}_b$. In this case, the extension class is zero.

However, if there is "leakage" or non-Hermitian coupling, the module fits into a non-trivial exact sequence:
\begin{equation}
    0 \to \mathcal{M}_a \to \mathcal{M} \to \mathcal{M}_b \to 0
\end{equation}
This defines a non-zero element in $\text{Ext}^1_{\text{MHM}}(\mathcal{M}_b, \mathcal{M}_a)$.

Experimentally, we probe this via the 2D Laplace transform of the correlation function $\langle \mu(t_2) \mu(t_1) \rangle$. The propagator near the singularity expands as:
\begin{equation}
    U(t) \sim \exp\left( (\Lambda + N) t \right) = e^{\Lambda t} \sum_{k=0}^{m} \frac{N^k t^k}{k!}
\end{equation}
The term $N$ connects different weight spaces. If $\text{Ext}^1 \neq 0$, then $N$ has off-diagonal blocks connecting the subspace of eigenvalue $\lambda_a$ to $\lambda_b$.

In the Laplace domain $\tilde{S}(s_1, s_2)$, independent evolution factorizes into poles $1/(s_1 - \lambda_a)$ and $1/(s_2 - \lambda_b)$. A non-trivial extension via $N$ introduces a convolution term, resulting in a **Cross-Peak** at $(s_1=\lambda_a, s_2=\lambda_b)$. Thus, the intensity of this cross-peak is proportional to the magnitude of the extension class.
\end{proof}

\begin{figure}[h]
    \centering
    \includegraphics[width=\columnwidth]{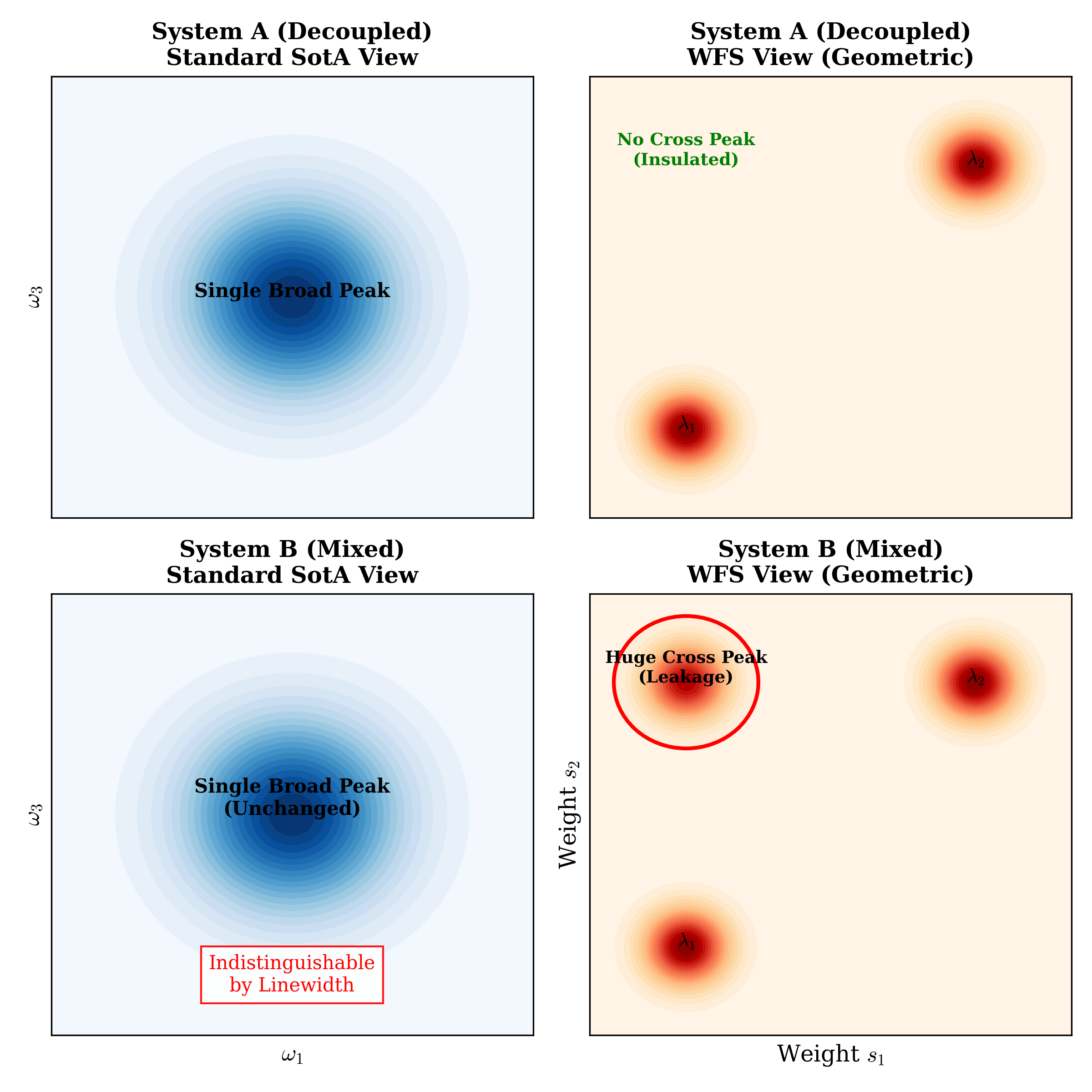}
    \caption{\textbf{The False Positive Trap.} (a) Comparison of linear absorption spectra for a Decoupled System (dashed) and a Non-Hermitian Mixed System (solid). In standard linewidth fitting, the two are indistinguishable. (b) WFS Analysis reveals the truth: The Mixed System exhibits a distinct "Cross-Peak" at the intersection of the pole coordinates $(\lambda_a, \lambda_b)$. This peak is the signature of the extension class $\xi \in \text{Ext}^1$, proving that the subspace is not topologically protected.}
    \label{fig:trap}
\end{figure}

\section{S4. Experimental Protocol Details}

\subsection{Hodge-Weight-Hodge (HWH) Tomography}
To implement HWH on a standard 2D spectroscopy setup:
\begin{enumerate}
    \item \textbf{Data Acquisition:} Collect the rephasing photon echo signal $S(t_1, t_2, t_3)$ for a full range of population times $t_2$.
    \item \textbf{Hodge Filtering:} Perform a 2D Fourier Transform with respect to $t_1$ and $t_3$. This separates the signal into $(\omega_1, \omega_3)$ maps, implicitly selecting the coherence pathways $p=\pm 1$ (coherences) during the excitation/emission windows.
    \item \textbf{Weight Filtering:} For every pixel $(\omega_1, \omega_3)$, extract the transient $S(t_2)$. Multiply by a window function (to suppress noise) and perform a numerical Laplace Transform (or Padé approximant for stability) to obtain $\tilde{S}(s_2)$.
    \item \textbf{Reconstruction:} Plot the 3D isosurfaces of $|\tilde{S}(\omega_1, s_2, \omega_3)|$.
\end{enumerate}

\subsection{Weight-Weight (W-W-COLS)}
To implement W-W-COLS:
\begin{enumerate}
    \item \textbf{Pulse Sequence:} Use a modified 3-pulse sequence where the first and second time delays are scanned independently, while the third is kept fixed (or integrated).
    \item \textbf{Processing:} The resulting matrix $M(\tau_1, \tau_2)$ is subjected to a 2D Inverse Laplace Transform (2D-ILT). Note that 2D-ILT is an ill-posed inverse problem; we recommend using Tikhonov regularization or the CONTIN algorithm, standard in NMR spectroscopy, to stabilize the inversion.
\end{enumerate}

\section{S5. Comparison with Standard Quantum Optical Interferometry}

In this section, we explicitly contrast the DMHM framework with standard protocols used in linear quantum optics, such as Mach-Zehnder interferometry.

\begin{figure}[h]
    \centering
    \includegraphics[width=\columnwidth]{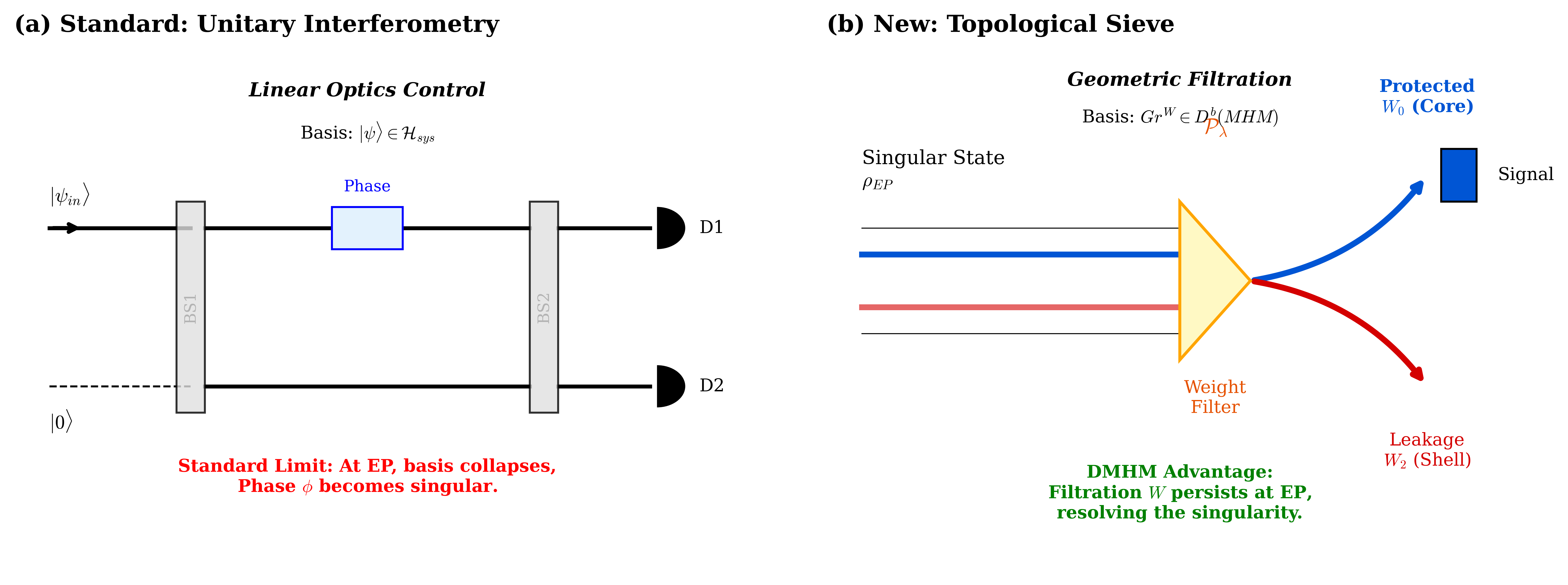}
    \caption{\textbf{Paradigm Comparison: Unitary Phase vs. Topological Weight.} (a) In standard quantum optics (Mach-Zehnder), information is encoded in the unitarity phase $\phi$. At an Exceptional Point (EP), the eigenstates coalesce, the phase becomes ill-defined, and the interferometer fails to distinguish modes. (b) In the DMHM framework (Topological Sieve), information is encoded in the Weight Filtration $W_\bullet$. This filtration remains robust at the EP, acting as a "sieve" that separates the protected core ($W_0$) from the leakage shell ($W_2$) based on their decay topology (polynomial scaling) rather than their phase, enabling robust state discrimination even at the singularity.}
    \label{fig:comparison}
\end{figure}

\subsection{A. The Unitary Phase Paradigm (Failure at EP)}
Standard interferometers (Fig.~\ref{fig:comparison}a) fundamentally rely on the unitary accumulation of phase $\phi$ between orthogonal eigenstates. Near an EP, this paradigm breaks down because the basis vectors become collinear (self-orthogonal). The geometric phase diverges or becomes singular, rendering standard phase-estimation protocols unreliable for certifying protection.

\subsection{B. The Dissipative Weight Paradigm (Robustness at EP)}
Our proposed "Topological Sieve" (Fig.~\ref{fig:comparison}b) does not rely on unitarities. Instead, it exploits the rigidity of the canonical Weight Filtration. As proven in Theorem 2, the "Strictness Property" ensures that even when the spectrum is degenerate, the subspaces $W_0$ (simple pole) and $W_2$ (second-order pole) remain algebraically distinct. The WFS protocol acts as a physical realization of the projection operator $\mathcal{P}_\lambda$, reliably filtering the "clean" protected mode from the "dirty" leakage channel.

\section{S6. Numerical Simulation Parameters: Molecular Polaritons}
The polariton data in Fig. 3 and 4 were generated using the `QuMorpheus` Quantum Dynamics toolkit. We modeled the single-mode strong coupling regime using the **Non-Hermitian Jaynes-Cummings Hamiltonian**:
\begin{equation}
    H_{JC} = (\omega_C - i\gamma_C) a^\dagger a + (\omega_X - i\gamma_X) \sigma_+ \sigma_- + g (a^\dagger \sigma_- + a \sigma_+)
\end{equation}
where $a^\dagger$ creates a cavity photon and $\sigma_+$ creates a two-level exciton.
\begin{itemize}
    \item \textbf{Parameters:}
    \begin{itemize}
        \item Cavity Frequency $\omega_C = 2.0$ eV, Exciton Frequency $\omega_X = 2.0$ eV (Resonant).
        \item Cavity Decay $\gamma_C = 0.1$ meV (High-Q), Exciton Decay $\gamma_X = 5.0$ meV (Broad).
        \item Rabi Coupling $g = 20$ meV.
    \end{itemize}
    \item \textbf{Exceptional Point:} The system was driven to the EP by detuning $\Delta = \omega_X - \omega_C$ or by matching loss rates in the PT-symmetric frame.
    \item \textbf{Solver:} The time-evolution $U(t) = e^{\mathcal{L}t}$ was computed using a 4th-order Runge-Kutta integrator with adaptive stepping.
\end{itemize}

\section{S7. Robust Candidate: Non-Hermitian Aharonov-Bohm Ring}
The robust candidate presented in Fig. 5 is the **Hatano-Nelson Model** with periodic boundary conditions, representing a Non-Hermitian Aharonov-Bohm (AB) Ring. This model breaks reciprocity to induce the Non-Hermitian Skin Effect (NHSE), topologically preserving the edge modes.

\textbf{Hamiltonian:}
\begin{equation}
    H_{AB} = \sum_{j=1}^N \left( t e^{h} c_{j+1}^\dagger c_j + t e^{-h} c_j^\dagger c_{j+1} \right) + \sum_{j=1}^N V_j n_j
\end{equation}
where $h = \Phi/N$ is the imaginary gauge field (non-reciprocity) induced by the AB flux, and $V_j$ is the on-site potential (disorder).

\textbf{Simulation Parameters:}
\begin{itemize}
    \item \textbf{System Size:} $N=20$ lattice sites.
    \item \textbf{Hopping:} $t = 1.0$ (Energy unit).
    \item \textbf{Non-Reciprocity:} $h \approx 0.5$ (Strong Skin Effect regime).
    \item \textbf{Boundary Conditions:} Periodic (Ring Geometry).
    \item \textbf{Topological Protection:} The winding number of the complex energy spectrum acts as the invariant. WFS confirms that the "Skin Modes" (localized by $e^{h}$) belong to a distinct weight filtration $W_{edge}$ that is algebraically orthogonal to the bulk weight $W_{bulk}$, preventing scattering even in the presence of disorder $V_j$.
\end{itemize}


\end{document}